# Coexistence of close packed structures in large substrate-free Ar-Kr clusters according to THEED data

O. G. Danylchenko, O. P. Konotop

B. Verkin Institute for Low Temperature Physics and Engineering of the National Academy of Sciences of Ukraine

47 Nauky Ave., Kharkiv, 61103, Ukraine

e-mail: *konotop@ilt.kharkov.ua*

A quantitative phase analysis of substrate-free single-component and binary clusters of the Ar–Kr system obtained by adiabatic expansion of gas into vacuum through a supersonic nozzle was performed. The studies were carried out *in-situ* using transmission electron diffraction technique (THEED) on clusters with an average size ranging from $2 \times 10^3$ to $1 \times 10^5$ atoms/cluster and across the entire range of component concentrations. The independence of the threshold size of clusters, corresponding to the beginning of the formation of the hcp phase, from the component composition was revealed. It was established that clusters larger than this threshold size have a two-phase fcc-hcp structure with an identical concentration of components in each phase. The fraction of the hexagonal phase increases with the size of the aggregations and depends on the component content, reaching maximum in clusters with an equimolar composition. Arguments are presented in favor of the formation of two-phase clusters in the supersonic jet, rather than separate single-phase fcc and hcp ones. These findings are in good agreement with the previously proposed thermally activated diffusion mechanism for the nucleation and growth of the hcp phase in rare gas clusters.

# 1. Introduction

Van der Waals forces are weak electrostatic interactions that occur between molecules due to the interaction of their dipole moments. A classic example of the application of Van der Waals forces in nature is geckos. Gecko feet are covered with microscopic hairs that use Van der Waals interactions to adhere to any smooth surface without releasing sticky liquid or using surface tension [1]. Van der Waals forces play a crucial role in various physical, chemical and biological processes, including the condensation of gases and the stabilization of molecular structures. The simplest examples of Van der Waals crystals are cryocrystals of rare gases. Rare gases have fully filled electron shells, condense at low temperatures (from 4.22 K for He to 211.5 K for Rn), and are well described by the Lennard-Jones pair potential; therefore, this class of substances is widely used for critical testing of theoretical models describing different properties of solids.

One of the most intriguing problems in the physics of solidified rare gases has long been the so-called fcc–hcp (face-centered cubic – hexagonal close-packed) dilemma [2], i.e. the problem of competition between two close-packed structures. Theoretical calculations of the crystal's potential energy give a slight advantage to hexagonal close-packing. However, in reality, the stabilization of the hcp structure in bulk crystals (except helium) is observed only under ultrahigh pressure conditions [3-6], and has a martensitic character, i.e. the transformation occurs through a collective shift of atoms of the close-packed plane from the fcc positions to the hcp sites. At the same time, the realization of the hcp phase under normal pressure conditions can occur in nanoscale and nanostructured systems. For example, a stable hcp phase has been observed in large (average size $\bar{N}$ greater than $10^4$ atoms per cluster) rare gas clusters [7-10]. Another example of the appearance of the hcp phase in nanoscale systems is in single-component [11, 12] and two-component [13] thin films of rare gases deposited on cooled substrates. However, while for thin-film materials, the temperature factor (annealing of samples [11, 12] or preliminary cooling of gas before condensation on the substrate [13]) plays a decisive role in the formation of the hexagonal phase, the transition from fcc to a two-phase structure in free clusters manifests with an increase in their size, i.e. it belongs purely to size effects. Such dependence of the structure and physicochemical properties of clusters on their size is the main feature of clusters. In fact, clusters are an intermediate state of matter between individual atoms or molecules and solid state, so studying them provides insights into how the properties of a solid emerge during the transition from an atom to a bulk material.

The object of research in this work is substrate-free clusters of the binary Ar–Kr system. The phase diagram for bulk Ar–Kr samples is of the cigar type, with only the fcc structure forming in the solid state [14, 15]. Results from electron diffraction studies of thin-film Ar–Kr samples [16-18], while showing discrepancies regarding the concentration intervals of component solubility, are consistent in that they show the formation of solid substitution solutions exclusively with the fcc structure. The hcp phase, specifically of almost pure argon rather than a mixture, was experimentally observed only in Ar-Kr cryocondensates deposited under special conditions involving pre-cooling of the gas mixture to a sub-liquid-nitrogen level [13]. As for clusters, we previously investigated the fcc-to-hcp transition in large ($\bar{N} \approx 2\times10^3 - 1\times10^5$ at./cl.) Ar–Kr clusters of equimolar component composition and found an intensification of hcp phase growth compared to pure Ar clusters [10]. The next task, which logically follows from the previous work is to investigate the competition between the two close-packed structures across the entire range of component concentrations. The solution of this task would allow us to determine the phase state diagram of Ar–Kr clusters and to confirm or correct our assumptions regarding the mechanism of nucleation and growth of the hcp phase in rare gas clusters.

## 2. Experiments

Substrate-free single-component and binary clusters of the Ar–Kr system were created by the method of adiabatic expansion of pre-cooled gas (or gas mixture) through a supersonic nozzle into a vacuum. This technique is well known for generating highly focused cluster beams, controlling the size and component composition of clusters, and is widely used in applied research to create nanostructured cluster-assembled materials [19], as well as cluster nanoplasma [20]. In this work, an original experimental complex was used, the main elements of which are a cluster beam generator, a cryostat with a built-in hydrogen cryopump, and an EMR-100M electron diffractometer. A detailed technical description of the setup can be found in [8]. Here, we will just outline the key experimental parameters. The conical nozzle for generating a supersonic jet had a critical cross-section diameter of 0.34 mm and a cone angle of 8.6º. The average size of the clusters in the jet was varied by adjusting the thermodynamic parameters of the gas mixture at the nozzle inlet: the pressure ($P_0$) ranged from 0.1 to 0.6 MPa; the temperature ($T_0$) ranged from 100 to 300 K. As a result, the average size of the produced clusters ($\bar{N}$) varied in the range of $2 \times 10^3 - 1 \times 10^5$ at./cl. For the binary Ar−Kr clusters, the krypton concentration ($C_{Kr}$) was maintained at 0.25±0.05 mf (mole fraction) and 0.8±0.05 mf, i.e. in the first case argon was the primary component, and in the second krypton was. A key factor for controlling the component composition of clusters is the necessity to take into account the enrichment effect in binary clusters. This effect causes the cluster composition to be enriched with the component that has a higher intermolecular binding energy, meaning a significant increase in krypton content within the Ar−Kr clusters compared to its content in the initial gas mixture [21]. To account for this factor, we constantly varied the required composition of the initial gas mixture according to the semiempirical equation previously determined [22] for the specific nozzle used in this work.

The transmission high-energy electron diffraction (THEED) method was used to determine the size, phase composition, and component composition of the clusters. The electron beam intersected the cluster beam at a distance of 100 mm from the nozzle inlet, where the clusters had already formed into stable aggregations. In order to reduce gas consumption, photographic registration of diffraction patterns with subsequent computer processing was used. A detailed description of the electron diffraction methods for determining the size and structure of clusters, as well as for performing quantitative phase analysis, can be found in [10]. The error in the experimental determination of the average linear cluster size $\delta$ did not exceed 10% ($\bar{N} \sim \delta^3$, i.e. the error in determining $\bar{N}$ is 30%). The error of quantitative phase analysis was up to 20% due to the overlap of the maxima of the fcc and hcp structures, as well as due to the significant contribution of the gas component and dimers to the diffraction spectrum.

## 3. Results and discussion

To determine the influence of the component composition of Ar–Kr clusters on their phase composition, experimental data from previous studies of single-component Ar clusters and binary Ar–Kr ones with an equimolar component content [10] were used, and new THEED experiments were performed for single-component Kr clusters as well as binary ones with krypton contents of 0.25 mf and 0.8 mf. Figure 1 shows diffraction pattern segments for Ar–Kr clusters with 0.25 mf krypton content, in the inverse lattice vector range from 1.5 to 2.5 Å$^{-1}$, which clearly illustrates the evolution of the phase composition of clusters as their size increases. In the case of clusters with an average size $\delta \approx 70$ Å ($\bar{N} \approx 5 \times 10^3$ at./cl.), the positions and intensities of the diffraction peaks indicate the formation of a substitution solid solution with an fcc structure (see Fig. 1(a)). A bad resolution of the fcc picks (111) and (200) is due to the presence of a large amount of stacking faults [10]. The diffraction pattern of clusters with a size of $\delta \approx 160$ Å ($\bar{N} \approx 5 \times 10^4$ at./cl.) shows a superposition of maxima from two structural phases: fcc and hcp (Fig. 1(b)). The positions of the peaks indicate that the concentrations of the chemical components in these phases are the same. Qualitatively similar patterns were observed for clusters with other krypton mole fractions, with only the integral intensities of the diffraction peaks changing. It should also be noted that the experimental conditions (temperature, pressure and component composition of the initial gas mixture) allowed the production of clusters whose size monotonically depended on the krypton content in the gas mixture, according to Hagena relation modified for binary clusters [23]. This means that the studied clusters showed neither the effect of heterogeneous intensification of Ar cluster growth on krypton nuclei [24] nor the inverse effect of cluster growth suppression by adding an excessively large amount of the impurity with a higher intermolecular binding energy to the initial gas [25].

Approximation of the experimental diffraction patterns with sums of Lorentzian functions allowed to measure the integrated intensities of the diffraction maxima for hcp (101) and fcc (200). These values were then used to determine the relative volumes of the crystalline phases, $V_{hcp}$ and $V_{fcc}$. Figure 2 shows the evolution of the ratio $V_{hcp}/V_{fcc}$ as the average cluster size increases, and its dependence on the component composition of the clusters. First and foremost, it was found that the threshold cluster size for the formation of the hexagonal phase is the same for both single-component and binary clusters, and is equal to $\delta \approx 90$ Å ($\overline{N} \approx 1 \times 10^4$ at./cl.). This result indicates that the initiation of hexagonal phase formation in Ar–Kr clusters is determined solely by the size of the aggregations and does not depend on their component composition. However, as the clusters further increase in size, their phase composition becomes sensitive to the concentration of the components. Firstly, the growth of the hexagonal phase fraction in two-component clusters is faster than in single-component argon or krypton ones. Secondly, in two-component clusters, the volume of the hexagonal phase depends on the component concentrations: the maximum fraction is observed at an equimolar mixture of argon and krypton. This result points to the dominant influence of static distortions in the crystal lattice caused by differences in the geometric sizes of the component molecules. At the same time, the temperature factor – namely, the higher the concentration of krypton in the formed clusters, the higher their temperature in the diffraction zone – is insignificant.

Fig. 2 also illustrates that the growth of the hcp phase fraction slows down with a further increase in cluster size. It is not possible to obtain single-phase aggregates with an exclusively hexagonal structure even at $\delta \approx 200$ Å ($\overline{N} \approx 1 \times 10^5$ at./cl.). This result is in good agreement with data for bulk rare gas cryocrystals under ultra-high pressure [6], which also indicate the impossibility of obtaining a single-phase hcp state. In order to understand how the absolute dimensions of the cubic phase evolve during the transition from single-phase to two-phase clusters, and why it is not possible to obtain clusters with only a hexagonal structure, the domain sizes of the two close packed phases were calculated as a function of the cluster size (Fig. 3). The black curves approximate the data for the fcc phase, and the red curves are for the hcp phase. The resulting dependencies have the same shape for both single-component (Fig. 3(a), 3(e)) and binary clusters (Fig. 3(b), 3(c), 3(d)). After the appearance of the hcp phase, the rapid growth of its size with a further increase in $\overline{N}$ shows a monotonous tendency to slow down. At the same time, the dependencies of the fcc domain sizes show a distinct bend in the range of $\overline{N}$ values from $6\times10^3$ to $3\times10^4$ at./cl. This interval is in good agreement with the cluster size range in which the number of stacking faults of the deformation type in the fcc lattice decreases to zero [10]. However, it is worth noting that while this $\overline{N}$ interval corresponds to a rapid increase in the size and fraction of the hcp phase, we do not observe a decrease in the size of the cubic domains, even in the case of an equimolar content of components, which is most favorable for the hexagonal structure, i.e. the fcc phase also continues to grow steadily. Such growth accelerates again within the mean cluster size range of more than $3\times10^4$ at./cl., and the forming fcc lattice is practically free of stacking faults. As a result, the two-phase fcc-hcp structure of large single-component Ar and Kr clusters reaches equilibrium and ceases to depend on their sizes. For binary clusters, it is reasonable to assume that the independence of their phase state from cluster sizes can be achieved at values of $\overline{N}$ exceeding $1\times10^5$ at./cl.

An important question for the investigation of fcc-to-hcp transition in rare gas clusters is whether they are actually two-phase aggregations, or separate single-phase fcc and hcp ones are formed in the cluster jet. For example, the authors of an X-ray study of the growth of argon clusters via particle fusion upon warming [26] suggest that in one part of the clusters a complete two-step structural transition of the fcc to intermediate orthorhombic phase to hcp occurs, while the other parts remain either with the fcc or orthorhombic structure. In contrast to the method used in [26] for obtaining clusters (namely, by injecting a jet of helium with a small admixture of Ar into superfluid helium followed by heating), in a supersonic gas jet, as it cools, the aggregates increase in size due to the coalescence of liquid droplets. Such method allows for the simultaneous growth of fcc and hcp structures from a liquid phase within a single aggregate. An additional argument for the two-phase composition of large free rare gas clusters comes from research of the size distribution of Ar clusters in a supersonic jet [27]. The distribution has a bell-shaped curve that shifts towards larger sizes as the initial gas pressure $P_0$ increases. This shift should inevitably lead to a situation where there are almost no clusters left in the jet with sizes less than $1 \times 10^4$ at./cl., which should to have a pure fcc structure. On the contrary, our experimental data indicate a constant increase in the absolute sizes of fcc domains, which is an indirect confirmation of the formation of two-phase fcc-hcp clusters.

## 4. Conclusions

A quantitative phase analysis was conducted on single-component and binary clusters of the Ar–Kr system using transmission high energy electron diffraction technique. The clusters were formed via supersonic adiabatic expansion of a gas jet in a vacuum. The studies were carried out in the cluster size range from $2 \times 10^3$ to $1 \times 10^5$ at./cl. and across the entire range of concentrations for both components.

It has been established that the threshold size of clusters, which corresponds to the beginning of the formation of the hcp phase, does not depend on the concentration of components in the clusters, and has the same value of approximately $1 \times 10^4$ at./cl. for both single-component and binary aggregates, With a further increase in cluster size, mixed two-phase fcc-hcp clusters are formed, in which the dependence of the phase fractions on the content of components is revealed. The fraction of the hexagonal phase in binary clusters is significantly larger than in single-component ones, with the greatest growth of the hexagonal structure observed at an equimolar content of the components. The preference of the hcp structure for binary clusters is associated with static distortions of the crystal lattice, which are known to facilitate diffusion processes.

It was shown that in clusters larger than $1 \times 10^4$ at./cl. the simultaneous growth of cubic and hexagonal structures is realized, which is an indirect confirmation of the formation of mixed two-phase fcc-hcp clusters in a supersonic jet, rather than single-phase fcc and hcp aggregates. It was found that the concentration of components in both close packed phases is the same. The simultaneous growth of close packed phases is also an additional argument in favor of the realization of a thermally activated diffusion process of nucleation and growth of the hcp phase in rare gas clusters. It is assumed that the process begins at the stage of crystallization of pre-condensed aggregates by forming the nuclei of a new structure on the basis of deformation-type stacking faults which are inherent for the fcc lattice in rare gas clusters.

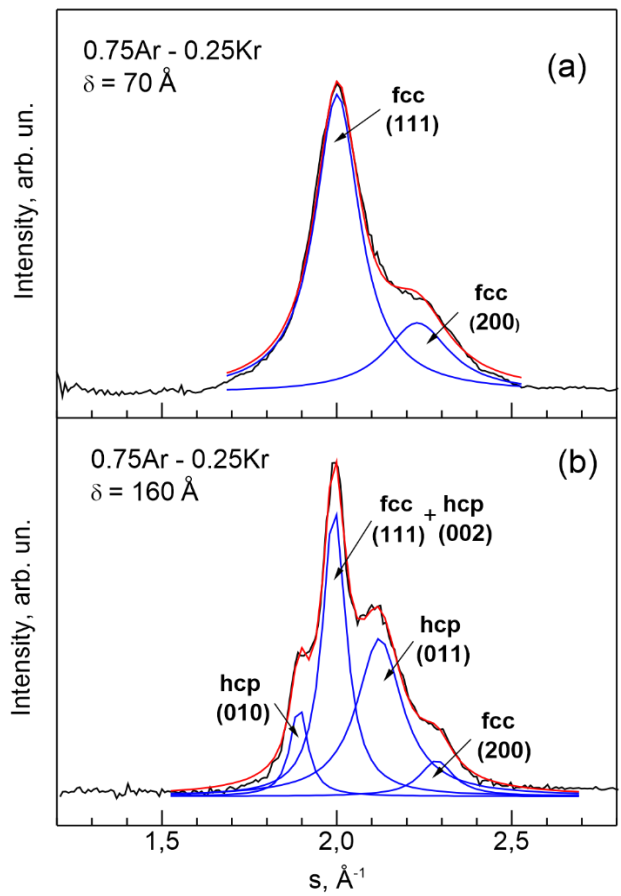

*Fig.* 1. Sections of electron diffraction patterns for binary clusters with the component molar fractions 0.75(Ar) – 0.25(Kr) illustrating the change in the phase composition when the average linear size of clusters $\delta$ changes from 70 Å (a) to 160 Å (b). Red and blue curves are approximations of experimental data by Lorentzian functions.

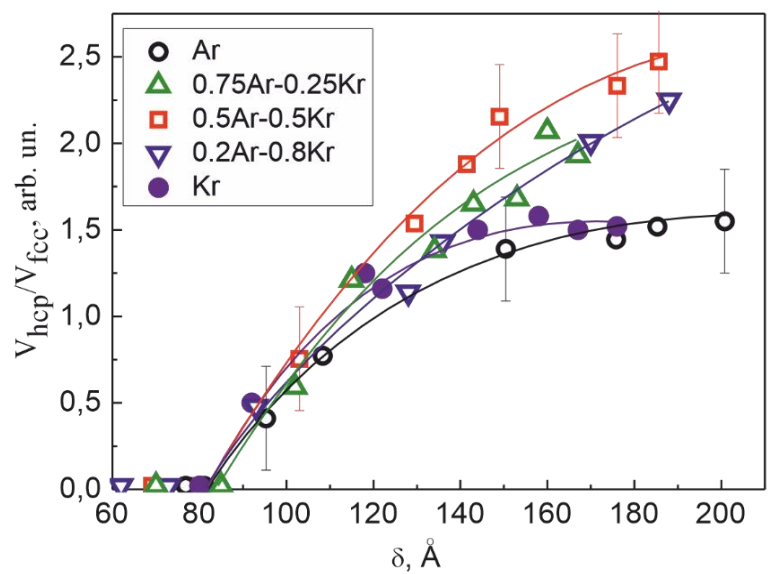

*Fig.* 2. Dependences of the ratio of the volume of the hcp phase to the volume of the fcc phase on the size of single-component and binary clusters of the Ar–Kr system.

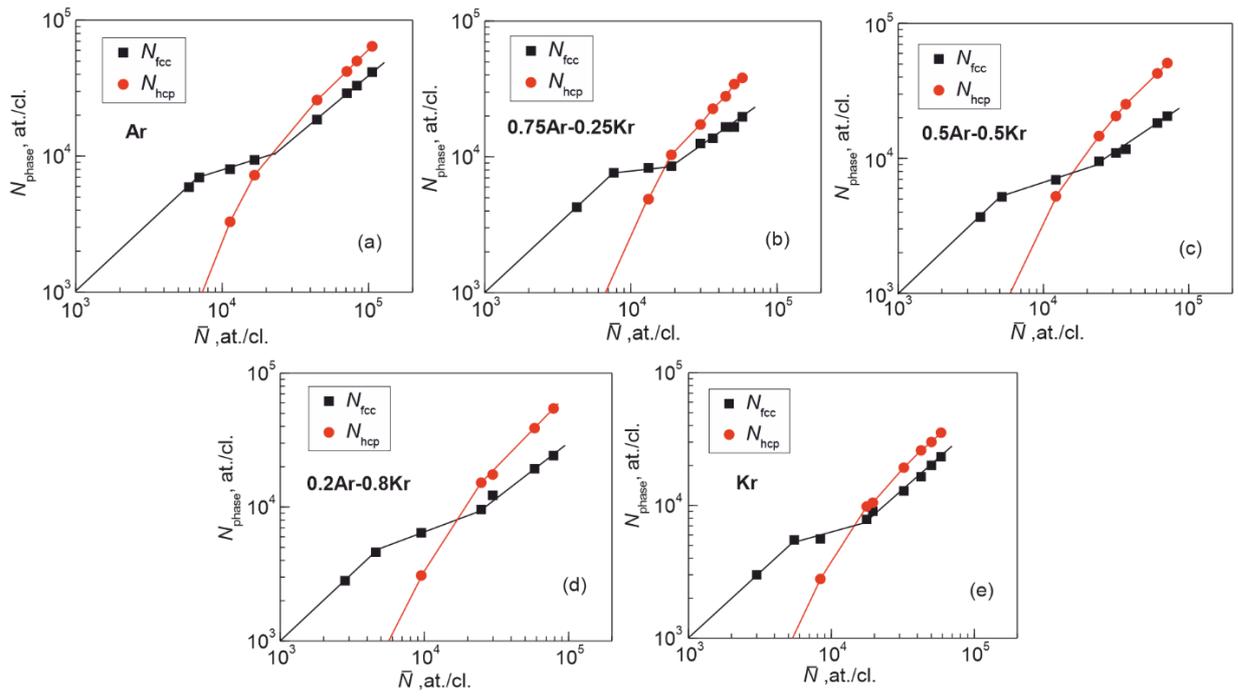

*Fig.* 3. Domain sizes of close packed phases in a cluster as a function of average cluster size. Black curves are approximations of data for the fcc phase, red curves are for the hcp phase.